\documentclass[pre,twocolumn,a4paper, amsfonts, amssymb, amsmath, reprint, showkeys, nofootinbib, twoside]{revtex4-1}
\usepackage{color,graphicx,hyperref,amsmath}
\usepackage{ulem}
\usepackage{mathtools}
\usepackage{graphicx}
\usepackage{float}

\begin{document}

\title{High frequency oscillations in spin-torque nano oscillator due to bilinear coupling}

\author{R. Arun}

\altaffiliation{  Department of Nonlinear Dynamics, School of Physics, Bharathidasan University, Tiruchirapalli-620024, India}

\author{R. Gopal}

\altaffiliation{ Department of Physics, Centre for Nonlinear Science and Engineering, School of Electrical and Electronics Engineering, SASTRA Deemed University, Thanjavur 613 401, India}

\author{ V. K. Chandrasekar}
\altaffiliation{ Department of Physics,	Centre for Nonlinear Science and Engineering, School of Electrical and Electronics Engineering, SASTRA Deemed University, Thanjavur 613 401, India}
\author{M. Lakshmanan}
\altaffiliation{  Department of Nonlinear Dynamics, School of Physics, Bharathidasan University, Tiruchirapalli-620024, India}



\begin{abstract}
Exchange coupling in an interfacial context is crucial for spin-torque nano oscillator (STNO) that consists of a non-magnetic spacer which is alloyed with a ferromagnetic material. Currently, investigations on the dynamics of the free layer magnetization and frequency enhancement in the STNO with bilinear coupling are still being actively pursued. In the present work, we investigate the dynamics of the STNO in the presence of bilinear coupling but in the absence of an external magnetic field by analyzing the associated Landau-Lifshitz-Gilbert-Sloncewski(LLGS) equation, and consequently the impact of the bilinear coupling on the dynamics of the magnetization of the free layer is studied.  It is observed that the frequency of the oscillations in the magnetization component along the direction of the pinned layer polarization can be enhanced above 300 GHz by positive bilinear coupling and up to around 30 GHz by negative bilinear coupling. We further reveal a transition from in-plane to out-of-plane precession both for positive and negative bi-linear couplings. We also analyze the switching of the magnetization for different values of current and bilinear coupling. Our detailed investigations of STNO with bilinear coupling aim at the possibilities of high-frequency devices by considering the applied current and bilinear coupling in the absence of a magnetic field. 
\end{abstract}
\maketitle
\section{Introduction}
A spin-polarized electrical current can impart spin angular momentum in the ferromagnetic material, which can be used to control the magnetization state of a magnetoresistive device called spin torque nano oscillator (STNO)~\cite{slon,berger,slon1,katine,deng,yama,liu,imai,wu,wolba,eklund,jiang,luba} .    In particular, it is feasible to cause the oscillations or precession of the magnetization, which is relevant for tunable microwave devices or to reverse the magnetization that is essential for various magnetic memory systems~\cite{groll}. In an STNO, two ferromagnetic layers are separated by a thin nonmagnetic, but conductive layer called a spacer. Among the two ferromagnetic layers, one is called the free layer, which is comparatively thinner than the other which is the pinned layer. In the free layer the direction of magnetization can change while it is fixed in the pinned layer. Further, some studies also ensure that the spacer layer can promote a high interlayer exchange coupling between its adjacent ferromagnetic layers~\cite{parkin}. The bottom and top layers of the two in-plane magnetized ferromagnetic layers are exchange-coupled via a Ruderman-Kittel-Kasuya-Yosida (RKKY) interaction across the thin nonmagnetic spacer, whose thickness is tuned to produce an antiferromagnetic coupling in zero applied field~\cite{parkin,ungu,naga,moser,gri}.

For instance, a nonmagnetic layer typically made of Ru~\cite{mck}  introduces a RKKY exchange coupling between two magnetic layers~\cite{mck}. The spin direction of the ferromagnetic layers can be parallel or antiparallel to each other depending upon the thickness of the spacer layer in magnetic multilayer systems. This parallel or antiparallel orientation of the ferromagnetic layers can be called collinear magnetization configuration~\cite{mck,duine}. On the other hand, obtaining a noncollinear magnetization configuration is possible due to the competition between the interlayer coupling energy and magnetic anisotropies of the coupled ferromagnetic layers for some structures. Recently, Nunn et al. have reported that the influence of the exchange coupling between two ferromagnetic layers (Fe) coupled through a nonmagnetic interlayer (Ru) is essential in controlling the magnetic layers' functionality~\cite{nunn}, and this has now been observed in various systems. It has been explained theoretically by several different approaches~\cite{cuc,rana,harris,grim,meena,tsu,arun,kuro,ishi}.

Recent results~\cite{nunn,ishi,kuro,jiang,luba,liu} in this context show that the presence of an exchange coupling system plays a backbone in the emergence of many spintronic-based applications such as magnetic field sensors, magnetic memory devices~\cite{rana, harris}, magnetic resistive random access memory (MRAM)~\cite{meena} and spin-torque nano oscillators~\cite{slon,slon1,berger}. Based on the nanoscale size and suitability for room-temperature operation, spin-torque oscillators (STOs) provide exciting possibilities for these applications. However, their adjustable range and oscillation frequency are only from 100 MHz to 10 GHz~\cite{grim,tsu}. Recently, we investigated and reported that the frequency of an STNO with bilinear and bi-quadratic couplings can be enhanced above 300 GHz by the current~\cite{arun}.  Also, Kurokawa et al. ~\cite{kuro} have shown the oscillations of the free layer magnetization in the components along the perpendicular directions of the pinned layer polarization with frequencies upto 576 GHz in the presence of bilinear and biquadratic interlayer exchange couplings in STNOs, and also with the free layer having low transition temperature for the saturation magnetization.  In their investigation they have shown that the biquadratic coupling is essential for the high frequency \cite{kuro}.

In this connection, our present report provides a detailed study on $\mathrm{Co \left|RuFe \right | Co}$ STNO with bilinear interlayer exchange coupling alone between the free and pinned ferromagnetic layers  and show the existence of oscillations of the free layer magnetization in the components along the pinned layer polarization with frequencies above 300 GHz with the free layer having high transition temperature. This unaccompanied role of the bilinear interlayer exchange coupling has been thoroughly researched since it has been used in many spintronics devices~\cite{ishi}, and multilayer magnetic thin films. Depending on the interfacial exchange coupling, both negative and positive exchange couplings have been seen in ferromagnetic/ferrimagnetic transition of metal and rare-earth alloy multilayer thin films~\cite{alt,mish} and the role of the bilinear coupling co-efficient are experimentally studied in Ref.~\cite{nunn}. However, numerical and analytical studies on the bilinear coupling in STNO without an external magnetic field that leads to magnetization oscillations have not been thoroughly studied in the literature~\cite {gusa}.

The paper is organized as follows. First, we formulate the model and the governing LLGS equation of motion and effective magnetic field for the present study in Sec. II. The positive and negative bilinear coupling dynamics and expression for minimum current for oscillations are presented in Sec. III and IV, respectively. Section V is devoted to the conclusion of the present work.

\section{Model}
\begin{figure}
	\centering\includegraphics[angle=0,width=0.5\linewidth]{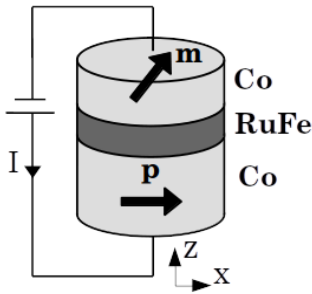}
	\caption{Schematic illustration of Co/ReFe/Co trilayer}
	\label{fig1}
\end{figure}
The schematic picture of an STNO considered for our study, which consists of a free layer, a spacer layer and a pinned layer, is shown in Fig.\ref{fig1}. The magnetization of the free layer is denoted as ${\bf M} = M_s{\bf m}$, where $M_s$ is the saturation of the magnetization. While the magnitude of the magnetization is fixed, its direction can change over time. The magnetization of the pinned layer ${\bf P} = M_s{\bf p}$ is fixed for both magnitude and direction. Here ${\bf m}$ and ${\bf p}$ are the unit vectors along ${\bf M}$ and ${\bf P}$, respectively.  As shown in Fig.\ref{fig1}, the positive and negative currents correspond to the flow of electrons from the free layer to pinned layer and vice versa, respectively. The free and pinned layers are considered to be made up of Co.   The spacer layer is a nonmagnetic conductive layer, constituting  an alloy of Ru and Fe.  The magnetization dynamics described by the LLGS equation that governs the motion of the unit vector  ${\bf m}$ is given as
\begin{align}
\frac{d{\bf m}}{dt}=&-\gamma{\bf m}\times{\bf H}_{eff}+ \alpha{\bf m}\times\frac{d{\bf m}}{dt} +\gamma H_{S}~ {\bf m}\times ({\bf m}\times{\bf p}).\label{llg} 
\end{align}
Here, $\gamma$  and  $\alpha$ are the gyromagnetic ratio and damping parameter, respectively. The spin-torque strength is 

\begin{align}
H_S = \frac{\hbar \eta I}{2 e M_s V (1+\lambda {\bf m}\cdot{\bf p}))},
\end{align}
where $\hbar$ is the reduced Planck's constant ($\hbar(=h/2\pi)$),    $I$ is the current, $e$ is the electron charge, and $V$ is the volume of the free layer, $\eta$ and $\lambda$ are the dimensionless parameters determining magnitude and angular dependence of the spin-transfer torque.
 
 The effective magnetic field $H_{eff}$ is given by 
\begin{align} 
 {\bf H}_{eff} = {\bf H}_{ani} + {\bf H}_{dem} + {\bf H}_{bil},
 \end{align}
where ${\bf H}_{ani}$ and ${\bf H}_{dem}$ is the anisotropy and the demagnetization field, respectively. The effective field also consists of a bilinear coupling interaction ${\bf H}_{bil}$ of interlayer exchange coupling between the free and reference layers, the details of which are given below. Specifically, the various interactions in (3) are given by
\begin{subequations}
\begin{equation}
{\bf H}_{ani} = H_k m_{z}~{\bf e}_z, \label{Han}
\end{equation}
\begin{equation}
{\bf H}_{dem} = -4\pi M_s m_{z}~{\bf e}_z, \label{Hde}
\end{equation}
\begin{equation}
{\bf H}_{bil} = -\frac{J}{dM_s}~{\bf e}_x.\label{Hbi}
\end{equation}
\end{subequations}
Consequently, we have
\begin{align}
{\bf H}_{eff}=(H_k-4\pi M_s) m_{z}~{\bf e}_z-\frac{J}{dM_s}~{\bf e}_x  \label{Heff}.
\end{align}
Here ${\bf e}_x$, ${\bf e}_y$ and ${\bf e}_z$ are the respective unit vectors along the positive $x$, $y$ and $z$ directions. $H_k$ is the magneto-crystalline anisotropy constant, $J$ is the coefficient of the bilinear coupling,  $M_{s}$ is the saturation magnetization and $d$ is the thickness of the free layer. The energy density of the free layer responsible for the effective field ${\bf H}_{eff}=- \partial E/\partial (M_s {\bf m})$  is given by
\begin{align}
E = &\frac{J}{d} ~{\bf m}.{\bf p} -\frac{M_s}{2}[ H_k -  4\pi M_s ]({\bf m}.{\bf e}_z)^2. \label{E}
\end{align}

The pinned layer is considered to be polarized along positive $x$-direction, i.e. ${\bf p} = {\bf e}_x$. The material parameters are adapted as $M_s$ = 1210 emu/c.c., $H_k$ = 3471 Oe, $\eta$ = 0.54, $\lambda$ = $\eta^2$, $d$ = 2 nm, $A$ = $\pi\times$60$\times$60 nm$^2$, $V$ = $Ad$, $\alpha$ = 0.005 and $\gamma$ = 17.64 Mrad/(Oe s). Since $H_k<4\pi M_s$, the system exhibits easy-plane anisotropy for $xy$-plane or hard axis anisotropy for $z$-axis due to the resultant demagnetization field $-(4\pi M_s-H_k) m_{z}~{\bf e}_z$. It means that the magnetization is always pulled towards the $ xy$ plane whenever it moves away from the plane with the strength directly proportional to $m_z$. Therefore, before applying any current, to minimize the energy (Eq.\eqref{E}), the magnetization of the free layer settles at (-1,0,0) for positive bilinear coupling ($J>0$) or (1,0,0) for negative bilinear coupling ($J<0$). This implies that the system exhibits antiferromagnetic coupling for the positive bilinear coupling and ferromagnetic coupling for the negative bilinear coupling between the free and pinned layers~\cite{mck}. It has been shown that the magnitude and sign of the bilinear coupling coefficient can be experimentally tuned by changing the concentration of Fe in the spacer layer made by Ru$_{100-x}$Fe$_x$ alloy~\cite{nunn} since the oscillations are observed when $I<0$ for the positive bilinear coupling and $I>0$ for the negative bilinear coupling, and both the cases of the bilinear couplings are investigated separately in the following sections.

\section{Dynamics for the positive bilinear coupling}

\begin{figure}
	\centering\includegraphics[angle=0,width=1\linewidth]{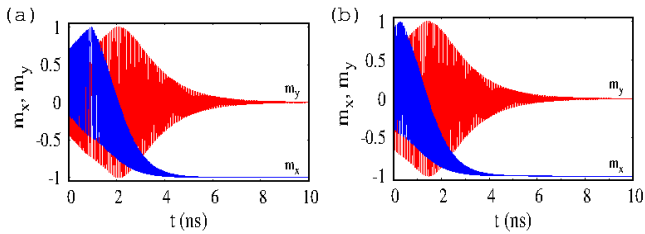}
	\caption{(a) Time evolution of $m_x$ and $m_y$ for (a) $J$ = 0.756 mJ/m$^2$ and (b) $J$ = 0.352 mJ/m$^2$.}
	\label{fig2}
\end{figure}

In the absence of current the equilibrium state of the unit magnetization vector ${\bf m}$ for the positive bilinear coupling  is ${\bf S}_1$ = (-1,0,0) since the field due to the interaction ${\bf H}_{bil}$ acts along the negative $x$-direction.  This is {\bf confirmed} in Figs.\ref{fig2}(a) and \ref{fig2}(b), where the time evolution of $m_x$ and $m_y$ are plotted for $J$ = 0.756 mJ/m$^2$ and 0.352 mJ/m$^2$, respectively, for different initial conditions. In both these figures \ref{fig2}(a) and \ref{fig2}(b), we can observe that the magnetization finally reaches the state ${\bf S}_1$. These numerical results coincide well with the experimental results obtained by Nunn $et~al$~\cite{nunn}, where the same system exhibits antiparallel configuration between the magnetizations of the free and pinned layers for $J$ = 0.756 mJ/m$^2$ and 0.352 mJ/m$^2$ corresponding to Ru$_{32}$Fe$_{68}$.

When the current is applied, depending upon the magnitude of the current, the system exhibits three different dynamics for ${\bf m}$. (i) When $|I|<|I_{min}|$, the unit magnetization vector ${\bf m}$ stays in the state ${\bf S}_1$ where it was existing already. (ii) When $|I_{min}|<|I|<|I_{max}|$, the vector ${\bf m}$ exhibits continuous precession. (iii)  When $|I|>|I_{max}|$ the vector ${\bf m}$ moves away from (-1,0,0) and settles into the state {\bf S}$_2$ (near (0,0,$\pm$1)) for small $J$ ($<$2.8 mJ/m$^2$) or settles into the state ${\bf S_3}$=(1,0,0) for large $J$ ($>$2.8 mJ/m$^2$). Hence the states ${\bf S}_1$, ${\bf S}_2$ and ${\bf S}_3$ are associated with the currents when $|I|<|I_{min}|$, $|I|>|I_{max}|$ for $J$ ($>$2.8 mJ/m$^2$) and $|I|>|I_{max}|$ for $J$ ($<$2.8 mJ/m$^2$), respectively. The critical value of the positive bilinear coupling strength $J_c$ = 2.8 mJ/m$^2$ is derived in Eq.\eqref{Jc}. Here, $I_{min}$ and $I_{max}$ are the minimum and maximum currents, respectively, between which oscillations can be exhibited.

\begin{figure*}
	\centering\includegraphics[angle=0,width=0.7\linewidth]{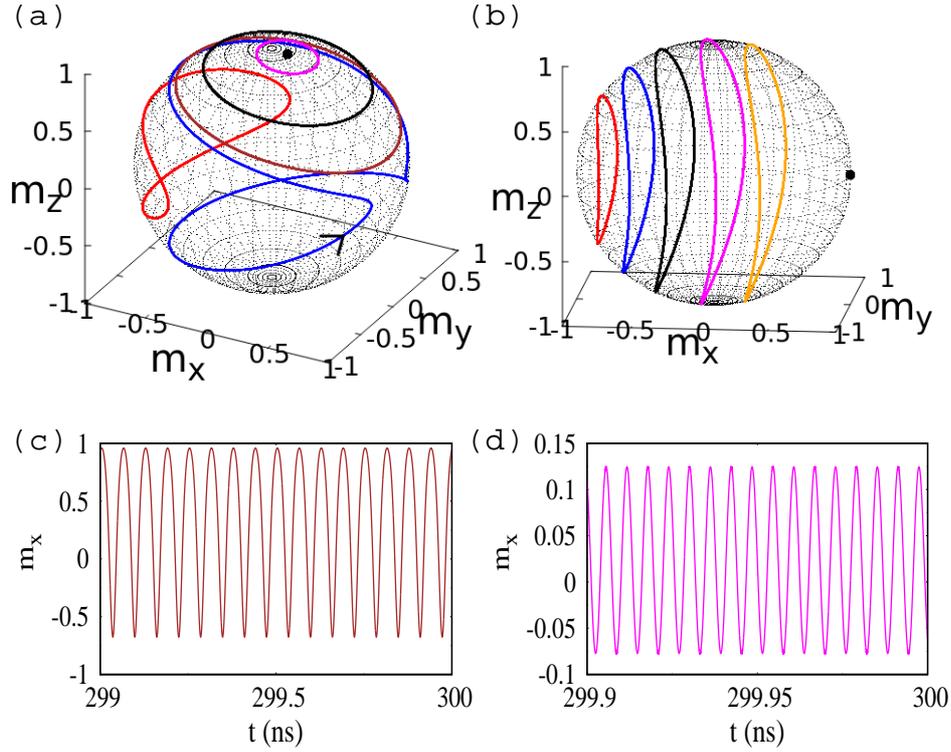}
	\caption{(a) Trajectory of ${\bf m}$ {\bf during $t$ = 299-300 ns} when $I$ = -0.5 mA (red), -1 mA (blue), -1.5 mA (brown), -2.5 mA \b(black), -3.25 mA (magenta) and -4 mA (black point) for $J$ = 0.4 mJ/m$^2$. (b) Trajectory of ${\bf m}$ {\bf during $t$ = 299-300 ns} when $I$ = -2 mA (red), -2.1 mA (blue), -2.2 mA (black), -2.3 mA (magenta), -2.35 (orange) and -3 mA (black point) for $J$ = 7 mJ/m$^2$.  (c) Time evolution of $m_x$ when $J$ = 0.4 mJ/m$^2$ and $I$ = -1.5 mA. (d) Time evolution of $m_x$ when $J$ = 7 mJ/m$^2$ and $I$ = -2.3 mA. }
	\label{fig3}
\end{figure*}

To confirm the precession of ${\bf m}$, oscillations of $m_x$ and tunability of the frequency by current, Eq.\eqref{llg} is numerically solved by adaptive step size Runge-Kutta-4 method.  The initial condition of ${\bf m}$, for the numerical simulation, is randomly chosen near the state ${\bf S}_1$.  When a negative current is applied with the magnitude $|I_{min}|<|I|<|I_{max}|$, the magnetization which was in the ${\bf S}_1$ state moves away from it due to the spin-transfer torque. This is due to the fact that the incoming electrons in the free layer, which were spin polarized along the positive $x$-direction, always move the magnetization to align with the positive $x$-direction. Once the magnetization moves away from the state ${\bf S}_1$ by STT, continuous precession is achieved due to the balance between the damping {\bf (due to the effective field)} and the STT.   The trajectories of ${\bf m}$ (after transition and between $t$ = 299 ns and $t$ = 300 ns) in continuous precession at different currents for a low value of $J$(= 0.4 mJ/m$^2$) and the time evolution of $m_x$ corresponding to $J$ = 0.4 mJ/m$^2$ and $I$ = -1.5 mA are plotted  in Figs.\ref{fig3}(a) and (c), respectively.  Similarly, the trajectories of ${\bf m}$ in the same duration for a high value of $J$(= 7.0 mJ/m$^2$) and the time evolution of $m_x$ corresponding to $J$ = 7 mJ/m$^2$ and $I$ = -2.3 mA are plotted in Figs.\ref{fig3}(b) and (d), respectively. We can observe from Fig.\ref{fig3}(a) that the trajectory corresponding to the current $I$ = -0.5 mA (red) exhibits in-plane precession around the $x$-axis due to the field from positive bilinear coupling. The direction of the precession is clockwise as seen from the positive $x$-axis.  When the strength of the current is increased further to $I$ = -1 mA (blue), the trajectory of the magnetization slightly transforms as shown in Fig.\ref{fig3}(a).  It seems that the trajectory has been folded along the negative $x$-axis.  The magnetization gets close to the positive $x$-axis when it reaches the $xy$-plane. This is due to the fact that the resultant demagnetization field becomes weaker when the magnetization gets closer to the $xy$-plane. Therefore the STT, which always moves the ${\bf m}$ towards the positive $x$-axis, becomes stronger and moves the magnetization towards the positive $x$-axis as much as possible.  Once the magnetization crosses the $xy$-plane, the magnetization moves away from the positive $x$-axis.  This is due to the fact that the resultant demagnetization field rotates the magnetization from negative to positive $y$-axis in the northern hemisphere and from positive to negative $y$-axis in the southern hemisphere.  When the current is further increased to -1.5 mA (brown), the magnetization shows a transition from the in-plane precession to out-of-plane precession around the $z$-axis as shown in the Fig.\ref{fig3}(a).  This is because an increase of curent increases the magnitude of the STT and consequently the projection of ${\bf m}$ in the $xy$-plane crosses the positive $x$-axis before the ${\bf m}$ reaches the $xy$-plane. Therefore the bilinear exchange coupling field and the resultant demagnetization field along with the STT precess the magnetization within the northern hemisphere continuously.  The out-of-plane precessions may symmetrically take place in the southern or northern hemisphere. Further increment in the current to -2.5 mA (black) and -3.25 mA (magenta) makes the concentric trajectories of ${\bf m}$ around the  equilibrium magnetization state where the ${\bf m}$ settles when $|I|>|I_{max}|$, with $I_{max}$ = - 3.4 mA for $J$ = 0.4 mJ/m$^2$.  The black point in Fig.\ref{fig3}(a) corresponds to the equilibrium state  at which the unit vector ${\bf m}$ settles for $I$ = -4 mA when $J$ = 0.4 mJ/m$^2$. This equilibrium state can be identified as follows:
The LLGS equation given by Eq.\eqref{llg} is transformed into spherical polar coordinates using the transformation equations $m_x=\sin\theta\cos\phi,~m_y = \sin\theta\sin\phi,~m_z=\cos\theta$ as
\begin{align}
	\frac{d\theta}{dt} ~=~ & \frac{\gamma}{1+\alpha^2} \Bigg \{ -\frac{J}{dM_s}(\alpha\cos\theta\cos\phi-\sin\phi)\nonumber\\ &-\alpha (H_k-4\pi M_s) \sin\theta\cos\theta\nonumber\\ &- H_{S0}\frac{(\alpha\sin\phi+\cos\theta\cos\phi)}{(1+\lambda \sin\theta\cos\phi)} \Bigg \} = P(\theta,\phi),\label{spllg1}\\
	~\frac{d\phi}{dt} ~=~ & \frac{\gamma \csc\theta }{1+\alpha^2} \Bigg \{\frac{J}{dM_s}(\cos\theta\cos\phi+\alpha\sin\phi)\nonumber\\ &+ (H_k-4\pi M_s) \sin\theta\cos\theta\nonumber\\& + H_{S0}\frac{(\sin\phi-\alpha\cos\theta\cos\phi)}{(1+\lambda \sin\theta\cos\phi)}  \Bigg \} = Q(\theta,\phi).\label{spllg2}
\end{align}

Here, $\theta$ and $\phi$ are the polar and azimuthal angles, respectively, $H_{S0} = \hbar\eta I/2eM_sV$. The equilibrium state is obtained from the equations $P(\theta^*,\phi^*)=0$ and $Q(\theta^*,\phi^*)=0$, where $\phi^*$ is numerically observed as $\phi^*\approx 0$. This leads us to derive the relation 
\begin{align}
\sin\theta^* = J/(dM_s(4\pi M_s-H_k)). \label{S2}
\end{align}
Therefore, the equilibrium state ${\bf S}_2$ for ${\bf m}$ when $|I|>|I_{max}|$ is given by ${\bf S}_2\approx(\sin\theta^*,0,\pm\cos\theta^*)$, where $\sin\theta^*$ is as given above.

However, when the magnitude of the current is increased much further than $|I_{max}|$, the equilibrium state will slightly move away from the state ${\bf S}_2$ and if the magnitude of the current is extremely large ($|I|>>|I_{max}|$), i.e above $\sim$100 mA, then the magnetization will settle in the state ${\bf S}_3$ = (1,0,0). 
\begin{figure}
	\centering\includegraphics[angle=0,width=1\linewidth]{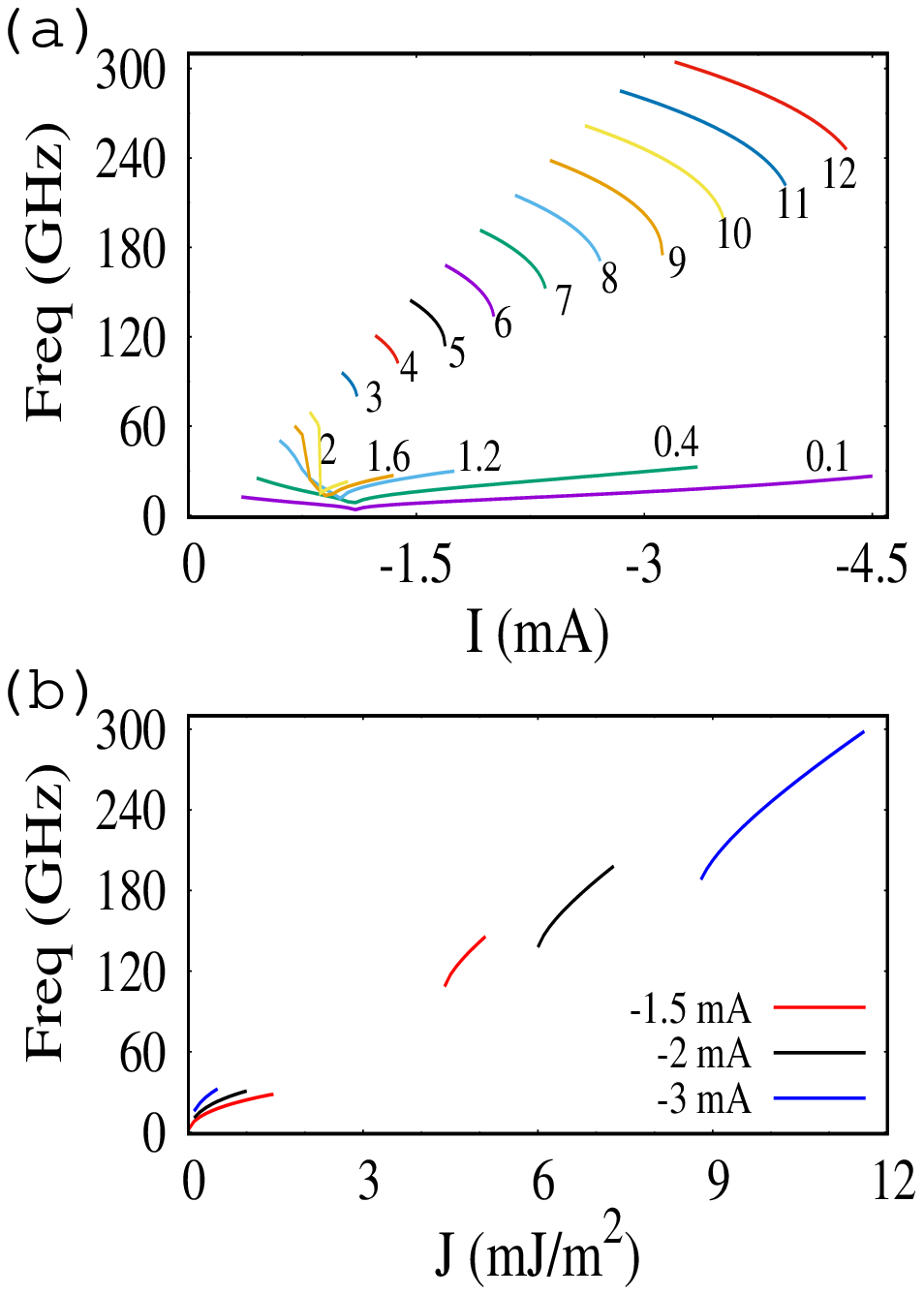}
	\caption{(a) Frequency tunability by current for different values of bilinear coupling $J$ (given in mJ/m$^2$) {\bf from 0.1 mJ/m$^2$ to 12 mJ/m$^2$}. (b) Frequency against bilinear coupling for different values of current $I$.}
	\label{fig4}
\end{figure}

From Eq.\eqref{S2}, we can understand that the value of $\theta^*$ becomes $\pi/2$ when $J = dM_s(4\pi M_s-H_k)$. It means that the equilibrium state ${\bf S}_2$ of the magnetization moves towards the state ${\bf S}_3$ = (1,0,0) as the strength of the positive bilinear coupling $J$ increases and reaches (1,0,0) when  $J\rightarrow J_c$, where 
\begin{align}
J_c = dM_s (4\pi M_s-H_k) = 2.8 ~{\rm mJ/m^2}.\label{Jc}
\end{align}

Similarly, the magnetization precession for the high strength of bilinear coupling ($J$ = 7.0 mJ/m$^2$) is also investigated by plotting the trajectories for the currents $I$ = -2 mA (red), -2.1 mA (blue), -2.2 mA (black), -2.3 mA (magenta), -2.35 mA (orange) and -3 mA (black point) in Fig.\ref{fig3}(b).    Unlike the case of low bilinear coupling as shown in Fig.\ref{fig3}(a), there is no transition from in-plane to out-of-plane precession while increasing the magnitude of the current and the magnetization exhibits in-plane precession only around the $x$-axis.   This can be reasoned as follows: When the strength of the bilinear coupling field is strong due to large $J(>0)$, the STT and the resultant demagnetization field are dominated by this bilinear coupling field.  Therefore, the rotations due to the resultant demagnetization field and the approach of the magnetization towards the positive $x$-axis due to the STT are not exhibited. When the current is increased further, the trajectory moves from the negative to positive $x$-axis and settles into the equilibrium state ${\bf S}_3$ when $|I|>|I_{max}|$, where $I_{max}$ = -2.35 mA for $J$ = 7.0 mJ/m$^2$.  The equilibrium state for the current -3 mA is shown by the black point in the Fig.\ref{fig3}(b).

To confirm the oscillations the time evolutions of the component $m_x$ are plotted  in Fig.\ref{fig3}(c) for $J$ = 0.4 mJ/m$^2$, $I$ = -1.5 mA and in Fig.\ref{fig3}(d) for $J$ = 7.0 mJ/m$^2$, $I$ = -2.3 mA.  The frequencies of the oscillations are 16 GHz and 163 GHz, respectively.

The frequencies of the oscillations, of $m_x$ are plotted against the current for different values of bilinear coupling strengths (given in mJ/m$^2$) {\bf from 0.1 mJ/m$^2$ to 12 mJ/m$^2$} in Fig.\ref{fig4}(a) and against bilinear coupling for different values of current in Fig.\ref{fig4}(b). From Fig.\ref{fig4}(a), we can understand that when the bilinear coupling coefficient is low, the frequency decreases up to some critical current $I_c$ and then increases. This change in the frequency from decrement to increment is attributed to the transition of magnetization precession from the in-plane to out-of-plane as discussed earlier with reference to Fig.\ref{fig3}(a). In Fig.\ref{fig4}(a), the existence of $I_{min}$ and $I_{max}$ is evident, and the range of current for the oscillations ($|I_{max}|-|I_{min}|$) confirms the wide frequency tunability by the current. The magnitude of $I_c$  slightly decreases with the increase of $J$. Also, we can observe that when $J$ is large ($\geq$2.9 mJ/m$^2$) the frequency decreases with the increase in the magnitude of the current up to $I_{max}$ and the $I_c$ does not exist. This is due to the nonexistence of out-of-plane precession, as shown in Fig.\ref{fig3}(b).  From Fig.\ref{fig4}(a) it is observed that the tunability range ($|I_{max}|-|I_{min}|$) decreases and increases with $J$ when the strength of $J$ is small and large, respectively.   At a given current, the frequency increases with the magnitude of bilinear coupling. Also, it is confirmed that the frequency can be enhanced up to 300 GHz for $J$ = 12.0 mJ/m$^2$ and even above when $J$ is increased further.

Similarly, the frequency is plotted against $J$ for different values of the current in Fig.\ref{fig4}(b).  Due to the nonexistence of out-of-plane precession at large strengths of $J$, the discontinuity appears in the frequency while increasing the value of $J$ as shown in Fig.\ref{fig4}(b).  From Fig.\ref{fig4}(b) we can observe that the frequency almost linearly enhances with $J$.  The frequency range is around 30 GHz and 300 GHz when the values of $J$ are small and large, respectively. The enlargement of frequency and switching time can be essentially attributed to the large value of the bilinear coupling strength $J$, which causes the system to behave more like a layered antiferromagnet~\cite{vol,vigo,roy,fira,khym}. The large value of $J$ in our system is possibly due to Nunn et al.'s recently proposed RuFe spacer layer~\cite{nunn}.  The current density corresponding to the frequency 299.6 GHz when $I$ = -3.35 mA can be obtained as 2.96$\times 10^7$ A/cm$^2$ for the cross sectional area $A=\pi\times 60\times 60$ nm$^2$.  Also, it is visible that the magnitude of the current can increase the range of $J$ for which the oscillations are possible.

\begin{figure}
	\centering\includegraphics[angle=0,width=1\linewidth]{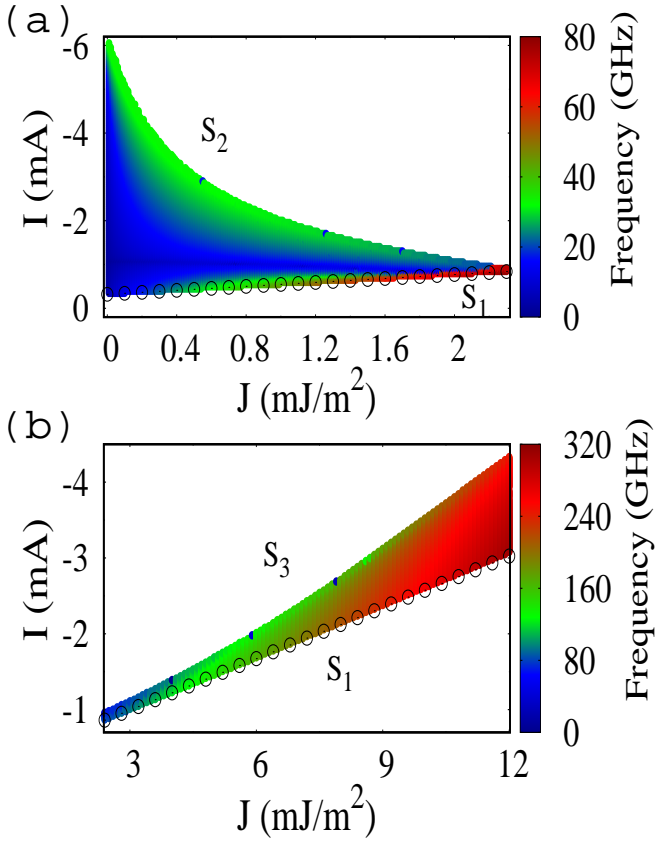}
	\caption{Frequency dependence on current and different ranges of bilinear coupling coefficient. {\bf The open circles are the minimum critical current $I_{max}$, for the onset of the oscillations, obtained from Eq.\eqref{Imin_pJ}.  {\bf S}$_1$ = (-1,0,0), {\bf S}$_2$ = $(\sin\theta*,0,\pm\cos\theta*)$ and {\bf S}$_3$ = (1,0,0) are the equilibrium states.}}
	\label{fig5}
\end{figure}

Figs.\ref{fig5}(a) and (b) summarize the dependence of the frequency on current and $J$ while $J$ is below and above 2.3 mJ/m$^2$, respectively. The white color region is nonoscillatory region. From Figs.\ref{fig5}(a) \& (b), we can see that the magnitude of the current above which the oscillations occur ($|I_{min}|$) linearly increases with $J$. The value $I_{min}$ for $J>$0 can be derived as follows:

The nature of the stability of an equilibrium state which is represented by polar coordinates can be identified from the following Jacobian matrix by using Eqs.\eqref{spllg1} and \eqref{spllg2}
\begin{align}
	\mathcal{J} = 
	\begin{pmatrix}
		{\left.\frac{dP}{d\theta}\right\vert}_{(\theta^*,\phi^*)} & {\left.\frac{dP}{d\phi}\right\vert}_{(\theta^*,\phi^*)}\\ {\left.\frac{dQ}{d\theta}\right\vert}_{(\theta^*,\phi^*)} & {\left.\frac{dQ}{d\phi}\right\vert}_{(\theta^*,\phi^*)}
	\end{pmatrix}. \label{Jmatrix}
\end{align}
The equilibrium state $(\theta^*,\phi^*)$ will be stable only when the system is dissipative about it. It will be dissipative if and only if the trace of the matrix $\mathcal{J}$ becomes negative,
\begin{align}
{\rm Tr}(\mathcal{J})<0.\label{Jcond}
\end{align}
We knew that when $|I|<|I_c^{min}|$ and $J>0$ the magnetization settles at ${\bf S}_1$, i.e, ($\pi/2,\pi$) in polar coordinates.  Therefore specific set of values $(\theta^*,\phi^*)=(\pi/2,\pi)$ satisfies Eq.\eqref{Jcond}. The trace of the matrix corresponding to ($\pi/2,\pi$) is given by
\begin{align}
{{\rm Tr}(\mathcal{J})|}_{(\theta^*,\phi^*)} = \frac{\gamma}{1+\alpha^2}\left[-\frac{2J\alpha}{dM_s}+(H_k-4\pi M_s)\alpha-\frac{2H_{S0}}{1+\lambda} \right].\label{stab1}
\end{align}
The minimum critical current $I_{min}$ (for $J>0$), below which the ${\bf S}_1$ is stable can be derived from Eqs.\eqref{Jcond} and \eqref{stab1} as
\begin{align}
I_{min} = \frac{eA\alpha(\lambda-1)}{d\hbar\eta}\left[2J+(4\pi M_s-H_k)dM_s\right]\label{Imin_pJ}
\end{align}
and it has been plotted as open circles in Figs.\ref{fig5}(a) and (b), which matches well with the numerical results and confirms the validity of the numerical results.  From Fig.\ref{fig5}(a) and (b) we can observe that value of $I_{max}$ decreases with $J$ at lower strengths of $J$ and increases (almost linearly) with $J$ at higher strengths of it.  Fig.\ref{fig5}(b) evidences that the range of current which exhibits oscillations increases with $J$ while $J$ is large.   In the case of positive current, the STT always moves the magnetization to be aligned with the negative x-direction. Therefore the positive current does not move the magnetization from the state (-1,0,0), where it existed already before the application of the current, and therefore no precession is exhibited.
\begin{figure}
	\centering\includegraphics[angle=0,width=1\linewidth]{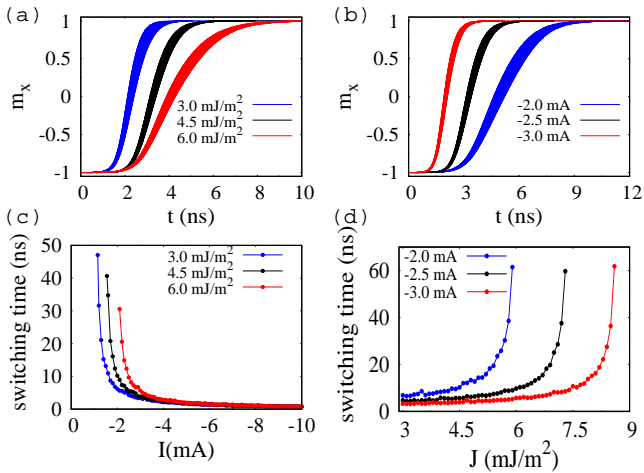}
	\caption{ Magnetization switching for different values of (a) $J$ when $I$ = -2.5 mA and (b) $I$ when $J$ = 4.5 mJ/m$^2$. Time to switch from (-1,0,0) to (1,0,0) with respect to (c) current and (d) bilinear coupling.}
	\label{fig5a}
\end{figure}
 We can observe in Figs.\ref{fig5}(a) and \ref{fig5}(b) that the magnetization settles into the equilibrium states {\bf S}$_2$ and {\bf S}$_3$, respectively when $I>I_{max}$.  It indicates a transition from {\bf S}$_2$ to {\bf S}$_3$ while increasing the strength of the positive bilinear coupling. As discussed in Eq.\eqref{Jc}, the transition occurs at $J$ = 2.8 ~${\rm mJ/m^2}$.

From Fig.\ref{fig5}(b), we can observe that when the magnitude of the current is above the magnitude of $I_{max}$, the magnetization will settle into the state ${\bf S}_3$ from ${\bf S}_1$ for the positive bilinear coupling. This indicates the existence of current-induced magnetization switching from the negative to positive $x$-direction. The corresponding switchings of $m_x$ from -1 to +1 for different values of bilinear coupling when $I$ = -2.5 mA and current when $J$ = 4.5 mJ/m$^2$ are plotted in Figs.\ref{fig5a}(a) and (b), respectively.   From Fig.\ref{fig5a}(a) we can observe that the switching times for $J$ = 3.0, 4.5 and 6.0 mJ/m$^2$ are 4.42, 6.01 and 9.42 ns, respectively.  Hence, the switching time increases with the magnitude of the positive bilinear coupling. On the other hand,  from Fig.\ref{fig5a}(b) we can understand that the switching times for the currents $I$ = -2.0, -2.5 and -3.0 mA are 9.88, 6.01 and 3.892 ns, respectively.  This implies that the switching times reduce with the increase of the magnitude of the current.  The variation of the switching time against current and the strength of the bilinear coupling for different values of $J$ and $I$ are plotted in Figs.\ref{fig5a}(c) and (d), respectively. Figs.\ref{fig5a}(c) and (d) confirm the decrement and increment of the switching time with the increase in the magnitude of current and positive bilinear coupling, respectively. Since the field due to the positive bilinear coupling acts along the negative $x$-direction, the enhancement in the magnitude of the negative current can quickly reverse the magnetization from negative to positive $x$-direction as shown in Fig.\ref{fig5a}(c). Similarly, when the strength of the positive bilinear coupling increases, its corresponding field along the negative $x$-direction increases, and consequently the magnetization takes much time to reverse from the negative to positive $x$-direction by the application of negative current as confirmed in Fig.\ref{fig5a}(d).   The above current-induced magnetization switching has spin torque magnetic random access memory applications and is much more efficient than the field-induced switching. The field-free switching may help produce magnetic memory devices with low power consumption and greater device density~ \cite{Loca, Ral}.

As observed from Figs.\ref{fig5}, when the current $I$ is kept constant and the strength of the positive bilinear coupling $J$ is increased, the magnetization reaches the equilibrium state ${\bf S}_2$ via out-of-plane precession (see Fig.\ref{fig3}(a)). When $J$ is increased further, the equilibrium state of the magnetization ${\bf S}_2$ becomes (1,0,0)  as $J\rightarrow J_c$ (see Eq.\ref{Jc} in the revised manuscript).  After the magnetization reaches the state ${\bf S}_3$ it continues to settle there without showing any oscillations until the further increase in $J$ is strong enough to move away the magnetization from the state ${\bf S}_3$ against the STT due to the incoming spin polarized electrons. As observed in Fig.\ref{fig4}(b) and Figs.\ref{fig5}, the gap between the offset of oscillations of ${\bf m}$ when reaching ${\bf S}_2$ and the onset of oscillations when emanating from ${\bf S}_3$ increases with the magnitude of the current. This is due to the fact that the strength of the STT which tends to keep the magnetization along the positive $x$-direction increases with the magnitude of current and consequently the strength of the bilinear coupling is required to be high enough to regain the oscillations from the equilibrium state ${\bf S}_3$.
\section{Dynamics for the negative bilinear coupling}

\begin{figure}
	\centering\includegraphics[angle=0,width=1\linewidth]{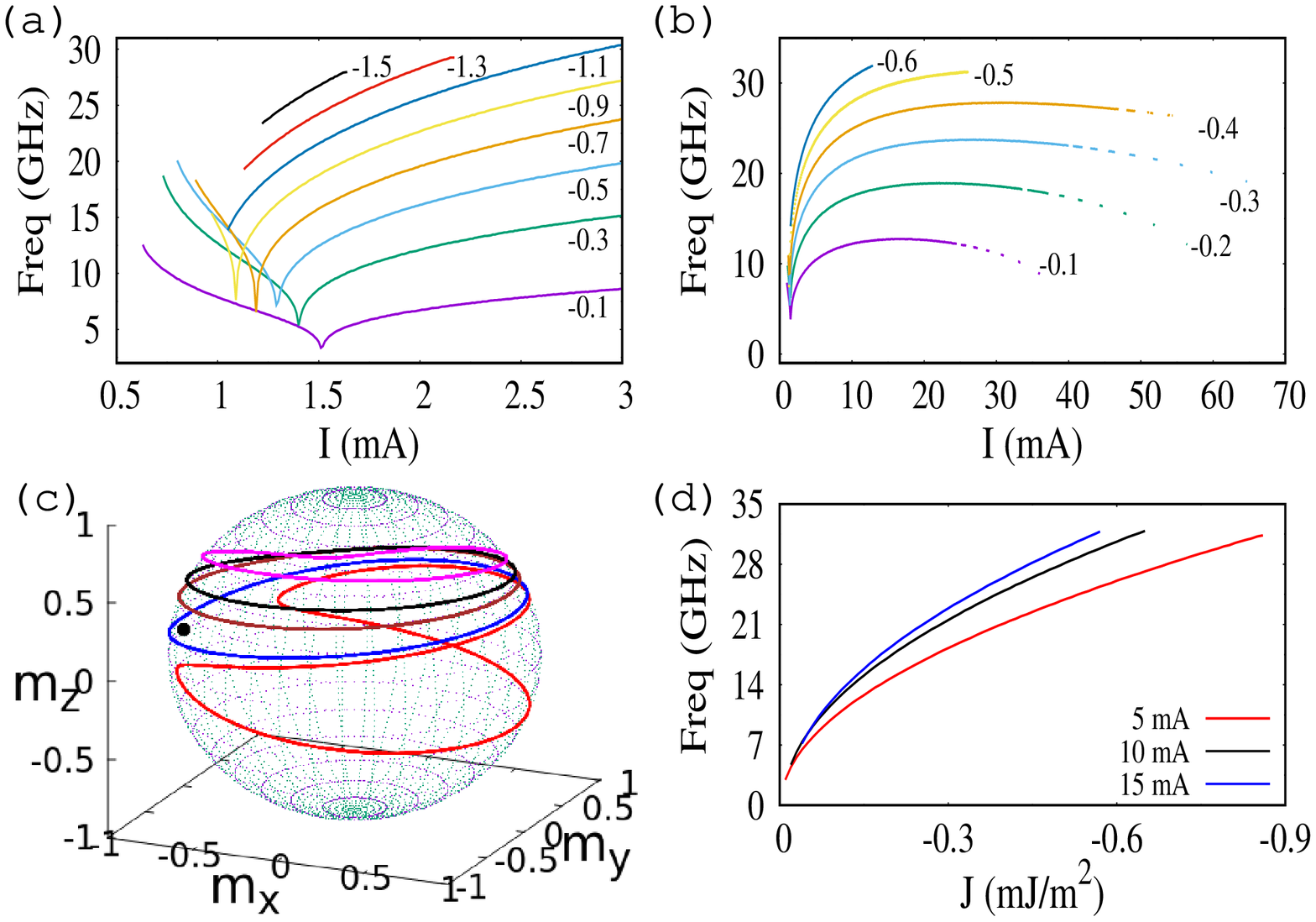}
	\caption{(a-b) Frequency tunability by current for different values of negative bilinear coupling $J$. (c) The magnetization trajectory when $I$ = 1 mA (red), 2 mA (blue), 10 mA (brown), 20 mA (black), 36 mA (magenta) and 37 mA (black point) for $J$ = -0.1 mJ/m$^2$. (d) The frequency variation against negative bilinear coupling for different values of current.}
	\label{fig6}
\end{figure}

In the presence of negative bilinear coupling the magnetization will initially be oriented at ${\bf S}_3$ since the field due to the negative bilinear coupling ${\bf H}_{bil}$ acts along the positive $x$-direction. The magnetization continues to be settled at ${\bf S}_3$ until the current  $I$ is increased to $I_{min}$. The STT, due to the positive current, will always move the magnetization to be aligned with the negative $x$-direction. When $I>I_{min}$, the magnetization is moved away from ${\bf S}_3$, and the system shows continuous precession for the vector ${\bf m}$.   The frequency of the oscillations of $m_x$ is plotted against low values of current in Fig.\ref{fig6}(a) and high values of current in Fig.\ref{fig6}(b) for different values of the negative bilinear coupling (given in mJ/m$^2$). From Fig.\ref{fig6}(a), we can understand that similar to the case of the positive bilinear coupling, the frequency decreases with current up to a critical value $I_c$ and then increases with current. Similar to the previous case, this increment in frequency after decrement is attributed to the transition from in-plane to out-of-plane precession. This is verified by plotting the trajectories of the vector ${\bf m}$ corresponding to $I$ = 1 mA (red) and 2 mA (blue) for $J$ = -0.1 mJ/m$^2$ in Fig.\ref{fig6}(c). Since the field, due to negative bilinear coupling, acts along the positive $x$-direction, the magnetisation trajectory corresponding to $I$ = 1 mA (red) has been folded along the positive $x$-axis and exhibits in-plane precession.

When the current increases to 2 mA (blue), the magnetization transforms from in-plane precession to out-of-plane precession in the northern hemisphere. However, the out-of-plane precession may also be symmetrically placed in the southern hemisphere.    The explanation behind this transition is similar to those discussed in the case of positive bilinear coupling. The out-of-plane precessions corresponding to the currents $I$ = 10 mA (brown), 20 mA (black) and 36 mA (magenta) for $J$ = -0.1 mJ/m$^2$ also are plotted in Fig.\ref{fig6}(c). From Fig.\ref{fig6}(a), we can understand that when the strength of the negative bilinear coupling is relatively high, the frequency shows  only an increment with the current. This is because at higher values of negative bilinear coupling, the unit magnetization vector ${\bf m}$  exhibits out-of-plane precession instead of exhibiting any transition from in-plane to out-of-plane precession.    In Fig.\ref{fig6}(b), the frequency is plotted up to large values of current for different values of $J$. The frequency increases with current and reaches its maximum. For small values of $J$, the frequency increases to its maximum and then decreases.   Fig.\ref{fig6}(b) shows that there is a maximum current $I_{max}$ above which oscillations are not possible. For the currents above $I_{max}$, the magnetization settles into ${\bf S}_1$ without showing any precession.

 In Fig.\ref{fig6}(b) we can observe the discontinuities for frequencies near $I_{max}$ upto $J\approx$ -0.4 mJ/m$^2$, where the system exhibits multistability i.e the magnetization may precess continuously or settle at ${\bf S}_1$. It is confirmed in Fig.\ref{fig6}(c) by precession for $I$ = 36 mA (magenta) and equilibrium state ${\bf S}_1$ for $I$ = 37 mA (black point).   In Fig.\ref{fig6}(b) it is observed that the discontinuities in the frequencies have disappeared above $J$ = -0.4 mJ/m$^2$. This is because the magnetization does not settle at ${\bf S}_1$ below $I_{max}$. 
 The magnetization exhibits three different nature of equilibrium states for $|J|>\sim0.4$ and  $I>I_{max}$.  When the current is increased near above $I_{max}$, the magnetization settles near poles at ${\bf S}_2$.  When $I$ is increased further the unit vector ${\bf m}$ settles into ${\bf S}_2$ or ${\bf S}_1$. If the current is increased further to extremely large values, the magnetization settles into ${\bf S}_1$.   The range of the current in which the oscillations are possible  ($I_{max}-I_{min}$) also increases (decreases) with $|J|$ when $|J|$ is small (large).

From Figs.\ref{fig6}(a) and (b), it is observed that the frequency can be reached around 30 GHz by increasing the current and the magnitude of the negative bilinear coupling.  In Fig.\ref{fig6}(d), the frequency is plotted against the negative bilinear coupling for different values of the currents.  It seems that the frequency increases almost linearly with the increase in the magnitude of negative bilinear coupling coefficient. Also, at a given $J$, the frequency increases with the magnitude of the current.

\begin{figure}
	\centering\includegraphics[angle=0,width=0.8\linewidth]{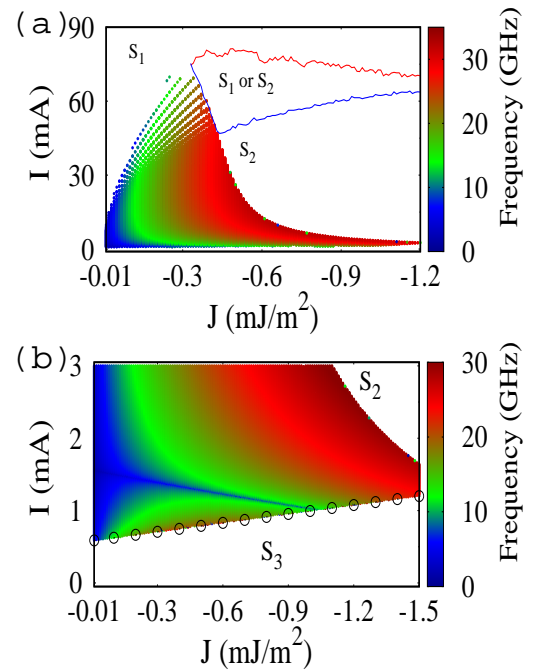}
	\caption{Dependence of the frequency on $J$ and $I$. (a) $I<$ 90 mA and (b) $I<$3 mA.}
	\label{fig7}
\end{figure}

The dependence of the frequency on the negative bilinear coupling and current is plotted for the large values of current in Fig.\ref{fig7}(a) and small values of current in Fig.\ref{fig7}(b). The white background corresponds to the non-oscillatory region. From Fig.\ref{fig7}(a) we can observe that the value of $I_{max}$ increases up to -0.33 mJ/m$^2$ and then decreases abruptly. From the bright green and red regions in Fig.\ref{fig7}(a) we can understand that the frequency can be maintained constant while increasing the current at fixed $J$. Also, it is clearly visible that the tunability  range of the frequency by current drastically reduces after $\sim$-0.3 mJ/m$^2$. This is different from the case of positive bilinear coupling where the ocillatory region ($|I_{max}|-|I_{min}|$) can be expanded with the increase of $J$. For currents above $I_{max}$, three different regions are identified for ${\bf m}$ as shown in Fig.\ref{fig7}(a). The three different regions for  equilibrium states ${\bf S}_1$, ${\bf S}_2$ and ${\bf S}_1$/${\bf S}_2$ for the current above $I_{max}$ are indicated in Fig.\ref{fig7}(a). To see the minute variation of frequency in the low current region, Fig.\ref{fig7}(b) is plotted for currents upto 3 mA. Fig.\ref{fig7}(b) confirms the decrement and increment in frequency with current when $|J|<1$ mJ/m$^2$. Also, the frequency at a given current increases with the strength of the negative bilinear coupling.

The minimum current $I_{min}$ for $J<$0 is similarly derived as in the previous case for positive bilinear coupling. When $I<I^{min}$  and $J<0$, the state ${\bf S}_3$ becomes stable and the magnetization settles into ${\bf S}_3$, corresponding to $(\pi/2,0)$ in polar coordinates. The trace of the matrix $J$ corresponding to the state $ (\pi/2,0)$ is derived as
\begin{align}
{{\rm Tr}(\mathcal{J})|}_{(\pi/2,0)} = \frac{\gamma}{1+\alpha^2}\left[\frac{2J\alpha}{dM_s}+(H_k-4\pi M_s)\alpha+\frac{2H_{S0}}{1-\lambda} \right]. \label{stab2}
\end{align}
From the condition \eqref{Jcond} and Eq.\eqref{stab2}, we can derive the minimum current (for $J<0$) below which the equilibrium state ${\bf S}_3$ is stable as
\begin{align}
I_{min} = -\frac{eA\alpha(1+\lambda)}{d\hbar\eta}\left[2J+(H_k-4\pi M_s)dM_s\right]\label{Imin_nJ}.
\end{align}
Eq.\eqref{Imin_nJ} is plotted in Fig.\ref{fig7}(b) as open circles and matches well with the numerical results. This confirms the validity of the numerical results.

If the current is negative, the STT always moves the magnetization towards the positive x-direction. Therefore the magnetization does not move from the state ${\bf S}_3$, where it was already existing before applying the current, by the negative current, and no precession is exhibited.
\begin{figure}
	\centering\includegraphics[angle=0,width=1\linewidth]{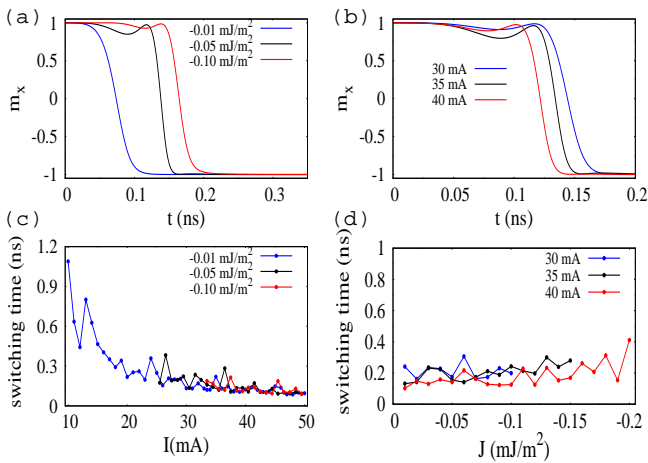}
		\caption{ Magnetization switching (negative bilinear coupling) for different values of (a) $J$ when $I$ = 33.5 mA and (b) $I$ when $J$ = -0.05 mJ/m$^2$. Time to switch from ${\bf S}_1$ to ${\bf S}_3$ with respect to (c) current and (d) bilinear coupling.}
	\label{fig7a}
\end{figure}

Similar to the case of positive bilinear coupling, magnetization switching can also be identified for negative bilinear coupling. As discussed in Fig.\ref{fig7}(a) when a current corresponding to the region of equilibrim state ${\bf S}_1$ is applied the magnetization will switch from ${\bf S}_3$ to ${\bf S}_1$. In Figs.\ref{fig7a}(a) and (b) the component $m_x$ is plotted to confirm the switching from positive to negative $x$-direction for different values of $J$ when $I$ = 33.5 mA and for different values of $I$ when $J$ = -0.05 mJ/m$^2$, respectively.  The variation of the switching time against current and the coupling is plotted in Figs.\ref{fig7a}(c) and (d), respectively. From Figs.\ref{fig7a}(a) and (c), we can understand that similar to the positive bilinear coupling, the switching time decreases with the increase in the magnitude of the current. Fig.\ref{fig7a}(d) confirms that there is no definite relationship between the switching time and the negative bilinear coupling. The switching time variation against the magnitude of the coupling is not smooth like in the case of positive bilinear coupling.

\section{Conclusion}
In conclusion, we have investigated the dynamics of   $\mathrm{Co \left|RuFe \right| Co}$ STNO using the LLGS equation and identified high-frequency oscillations in the magnetization of the free layer due to the presence of bilinear coupling. The obtained orientations of the magnetization of the free layer with that of the pinned layer in the absence of current match well with the experimental results. A transition in the precession of the magnetization from in-plane precession to out-of-plane precession while increasing the current is observed for both positive and negative bilinear coupling cases. However, the transition does not occur at higher strengths of the bilinear coupling. Only an in-plane precession for the positive bilinear coupling and an out-of-plane precession for the negative bilinear coupling are exhibited.   A wide range of frequency tunability by the current is observed for both cases of bilinear coupling. While the frequency is enhanced upto 30 GHz by the negative bilinear coupling, the positive bilinear coupling enhances the frequency upto and above 300 GHz.  This high frequency has been shown for the oscillations of the magnetization vector (free layer) along the pinned layer polarization and with the free layer having high transition temperature for the saturation magnetization.   The range of the current in which the frequency can be tuned increases with the strength of the positive bilinear coupling corresponding to the in-plane precession. 

Oscillations are exhibited for the positive (negative) bilinear coupling when the current is applied in the negative (positive) direction. Also, oscillations are possible only when the current is between $I_{min}$ and $I_{max}$. When $|I|<|I_{max}|$, the magnetization settles into (-1,0,0) for $J>0$ and (1,0,0) for $J<0$. If the strength of the positive bilinear coupling is large, then the magnetization settles into (1,0,0) for all the magnitudes of the current above $|I_{max}|$. On the other hand, if the strength is small, it settles near poles (${\bf S}_2$) when $|I|>|I_{max}|$ or into (1,0,0) when $|I|>>|I_{max}|$. If the bilinear coupling is negative, there are three regions corresponding to the  equilibrium states ${\bf S}_2$, ${\bf S}_1$ (or) ${\bf S}_2$ and ${\bf S}_1$  above $I_{max}$ depending upon the values of $I$ and $J$. The magnetization switching induced by the current alone is identified for both of the bilinear couplings. It is observed that the switching time reduces with the increase in the magnitude of the current for both cases of the bilinear coupling. 

We have also analyzed the expressions for the minimum currents to achieve the oscillations for both the positive and negative bilinear couplings. We have shown that they match well with the numerically obtained results. We have also proved that the bilinear coupling is sufficient for the high-frequency oscillations among two interlayer exchange couplings, namely bilinear and biquadratic couplings.  We wish to point out that this study has been carried out for the temperature $T$ = 0 K. However, the free layer we have considered is perpendicular magnetic anisotropic one and this is normally robust against thermal noise \cite{Tudu}.
We believe that our detailed study on bilinear coupling can be helpful in applications related to microwave generation with high-frequency enhancement and magnetic memory devices.

\section*{Acknowledgement}
The works of V.K.C. and R. G are supported by the DST-SERB-CRG Grant No. CRG/2020/004353 and they wish to thank DST, New Delhi for computational facilities under the DST-FIST programme (SR/FST/PS-1/2020/135) to the Department of Physics.   M.L. wishes to
thank the Department of Science and Technology for the award of a DST-SERB National Science Chair  under Grant No. NSC/2020/00029 in which R. Arun is supported by a Research Associateship.

\end{document}